\documentclass[twocolumn]{aastex631}
\usepackage{graphicx} 
\usepackage{graphics}
\usepackage{multirow}
\usepackage{subfigure}
\usepackage{placeins}
\usepackage{appendix}
\usepackage{tabularray}
\usepackage{color}
\usepackage{url}
\usepackage{rotating}

\shorttitle{Search EM counterparts to GW231109\_235456}
\shortauthors{Li et al.}

\begin{document}

\title{Systematic Search for Electromagnetic Counterparts to the Binary Neutron Star Merger Candidate GW231109\_235456}

\author[0000-0002-7598-9250]{Zhirui Li}
\affiliation{National Astronomical Observatories, Chinese Academy of Sciences, Beijing 100012, P.R. People’s Republic of China.}
\affiliation{School of Astronomy and Space Science, University of Chinese Academy of Sciences, Beijing 100049,  People's Republic of China.} 

\author[0000-0001-7952-7945]{Zhiwei Chen}
\affiliation{National Astronomical Observatories, Chinese Academy of Sciences, Beijing 100012, P.R. People’s Republic of China.}
\affiliation{School of Astronomy and Space Science, University of Chinese Academy of Sciences, Beijing 100049,  People's Republic of China.} 

\author[0000-0003-3250-2876]{Yang Huang}
\affiliation{School of Astronomy and Space Science, University of Chinese Academy of Sciences, Beijing 100049,  People's Republic of China.} 
\affiliation{National Astronomical Observatories, Chinese Academy of Sciences, Beijing 100012, P.R. People’s Republic of China.}

\author[0000-0002-1310-4664]{Youjun Lu}
\affiliation{National Astronomical Observatories, Chinese Academy of Sciences, Beijing 100012, P.R. People’s Republic of China.}
\affiliation{School of Astronomy and Space Science, University of Chinese Academy of Sciences, Beijing 100049,  People's Republic of China.} 

\author[0000-0002-2874-2706]{Jifeng Liu}
\affiliation{National Astronomical Observatories, Chinese Academy of Sciences, Beijing 100012, P.R. People’s Republic of China.}
\affiliation{School of Astronomy and Space Science, University of Chinese Academy of Sciences, Beijing 100049,  People's Republic of China.} 

\correspondingauthor{Zhiwei Chen, Yang Huang and Youjun Lu}
\email{chenzhiwei171@mails.ucas.ac.cn}
\email{huangyang@ucas.ac.cn }
\email{ luyj@nao.cas.cn}

\begin{abstract}
In this letter, we present a systematic search for the electromagnetic counterparts of binary neutron star (BNS) merger candidate GW231109\_235456 by examining all transients reported within the 90\% probability region and detected within four days of the merger. While non-detection in $\gamma$-ray, we identify two optical candidates, each associated with and residing in a host galaxy, which locate within 330~Mpc from earth; notably, one of them, AT2023xqy, is located at a distance of $178.6$\,Mpc, in good agreement with the estimated distance of the GW candidate ($\sim165^{+70}_{-69}~\mathrm{Mpc}$). Near the trigger time of GW231109\_235456 (MJD~60257.996), AT2023xqy showed evidence of a $\sim$15-day rise, first detected at $3\sigma$ significance on MJD~60259.097 and confirmed above $5\sigma$ on MJD~60262.088. This was followed by a rapid $\sim$5-day decline and a plateau lasting at least 50~days, with the subsequent decay unobserved due to a data gap. The spatiotemporal coincidences indicate that AT2023xqy could be a candidate for the EM counterpart of BNS merger candidate GW231109\_235456, though its lightcurve is difficult to reconcile with a standard kilonova. We examine two possible scenarios to explain the origin of AT2023xqy, a BNS merger-irrelevant scenario involving a peculiar supernova or a BNS merger-relevant scenario involving a magnetar-powered kilonova under extreme conditions. Follow-up radio observations are strongly encouraged, as they may provide critical insights into the nature of AT2023xqy.
\end{abstract}


\section{Introduction}
\label{sec:intro}

\begin{table*}[ht]
\centering
\caption{8 transients within 90\% probability area $<4$ days post-GW detection.}
\begin{tblr}{
  cells = {c},
  hline{1,10} = {-}{0.08em},
  hline{2} = {-}{},
}
Designation & {R.A. (J2000.0)\\Dec. (J2000.0)}        & {Date of Discovery\\MJD}       & {Mag. at Discovery \\ Filter}&{Host Galaxy\\Redshift $z$}                                     \\
AT2023abkc & {23:53:50.789\\$-$28:09:41.22} & {2023-11-10 05:32:09.024\\60258.23066} & {20.40\\PanSTARRS $w_\mathrm{P1}$}         & {PGC 744988\\0.1075}                                         \\
AT2023xqf  & {00:03:55.159\\$-$29:35:38.95} & {2023-11-10 10:40:28.128\\60258.44477} & {20.08\\GOTO \textit{L}}             & ---                                                          \\
AT2023xpx  & {23:38:49.625\\$-$29:40:58.84} & {2023-11-11 10:09:02.016\\60259.42294} & {19.17\\GOTO \textit{L}}             & {PGC 726409 (AGN)\\0.1026}                                   \\
AT2023xnj  & {00:21:31.492\\$-$32:48:20.18} & {2023-11-11 10:45:19.296\\60259.44814} & {20.25\\GOTO \textit{L}}             & {PGC 685271\\0.176$\pm$0.021 (photo-$z$)} \\
AT2023acmp & {00:14:18.685\\$-$30:23:38.80} & {2023-11-12 03:09:29.000\\60260.13159} & {26.02\\JWST other}               & {A2744cl 1447\\1.1613}                                           \\
AT2023xsz  & {00:45:48.771\\$-$28:48:25.65} & {2023-11-12 22:48:42.336\\60260.95049} & {19.114\\ATLAS \textit{c}}           & {PGC 737216\\0.070$\pm$0.013 (photo-$z$)} \\
AT2023xqy  & {23:41:43.058\\$-$34:12:06.46} & {2023-11-13 11:06:02.592\\60261.46252} & {19.32\\GOTO \textit{L}}             & {ESO 408-16 / PGC 72135\\0.0390}                             \\
AT2023xsy  & {23:35:35.668\\$-$40:12:40.81} & {2023-11-13 19:55:11.136\\60261.82999} & {19.49\\ATLAS \textit{c}}            & {PGC 2794577\\0.1273}                                        
\end{tblr}
\end{table*}

Binary neutron star (BNS) mergers are prime targets in multi-messenger astrophysics, as they generate both gravitational wave (GW) signals and electromagnetic (EM) counterparts such as kilonovae, short Gamma Ray bursts (sGRBs) and multi-band afterglows. The landmark event GW170817, detected by LIGO \citep{LIGOScientific:2014pky} and Virgo \citep{VIRGO:2014yos} over eight years ago, was accompanied by both the sGRB GRB170817A and kilonova AT2017gfo, with extensive follow-up observations across the entire EM spectrum \citep{LIGOScientific:2017vwq, LIGOScientific:2017ync, 2017ApJ...848L..12A}. These multi-messenger observations provided unprecedented insights into r-process nucleosynthesis \citep{2017Natur.551...80K}, the formation and structure of relativistic jets \citep{2018Natur.561..355M}, and the dynamics of compact object mergers \citep{2019ApJ...876..139G}. In contrast, the second reported BNS merger candidate, GW190425 \citep{LIGOScientific:2020aai}, lacked a confirmed EM counterpart, highlighting the difficulty of capturing such signals from these events.

The ongoing fourth observing run of the LIGO–Virgo–KAGRA (LVK) network, which began in May 2023, offers unprecedented sensitivity to GW events \citep{LIGO:2024kkz, Capote:2024rmo, LIGOO4Detector:2023wmz}. However, no confident BNS candidates have been reported in real-time searches or in the latest Gravitational Wave Transient Catalog \citep[GWTC-4.0;][]{gwtc4}. The LVK official pipelines are designed to detect not only BNS, but also binary black hole (BBH) and neutron star–black hole (NSBH) mergers. In such cases, weak BNS signals that would otherwise be consistent with realistic BNS populations can fall below detection thresholds due to the non-optimized population priors. These sub-threshold mergers may still produce observable EM counterparts, motivating the need for refined search strategies.

Recently, \citet{Niu25} proposed a search method optimized for sub-threshold GW events using the Galactic double neutron star (DNS) catalog, built from radio observations \citep{2016ARA&A..54..401O}, as a representative BNS population. Because this population is considerably narrower than the broad parameter space explored in O4a \citep{gwtc4}, this approach can enhance the recovery of BNS signals that might be missed by standard pipelines. Applying this refined search to the first part of LVK’s fourth observing run, \citet{Niu25} identified a significant trigger with a false-alarm rate of one per 50 years and a network signal-to-noise ratio of 9.7. This event was initially reported in LVK low-latency processing as S231109ci and later included in GWTC-4.0 as GW231109\_235456, a sub-threshold BNS candidate. If of astrophysical origin, the inferred source properties suggest component masses of $1.40$--$2.24~M_\odot$ for the primary and $0.97$--$1.49~M_\odot$ for the secondary, yielding a total mass of $2.95^{+0.38}_{-0.07}~M_\odot$. The event was localized to a $450~\mathrm{deg^2}$ region (90\% probability) at a luminosity distance of $165^{+70}_{-69}~\mathrm{Mpc}$.

In this letter, we present a systematic search for the EM counterparts of GW231109\_235456 using archival data. We identify two potential optical transients, with one, namely AT2023xqy lying well within both the sky localization map and the expected luminosity distance. Furthermore, its rising time is consistent with the GW trigger time, making it a plausible EM counterpart candidate to GW231109\_235456. The structure of this letter is as organized as follows. In Section~\ref{sec:seraching}, we describe the complete procedure for searching the EM counterpart; Section~\ref{sec:scenario} presents the light curves of the potential candidates and interprets them using a simple toy model. Finally, Section~\ref{sec:summary} summarizes our findings. Throughout this paper, we adopt the cosmological parameters from the Planck 2018 results \citep{2020A&A...641A...6P}.

\begin{figure*}
\plotone{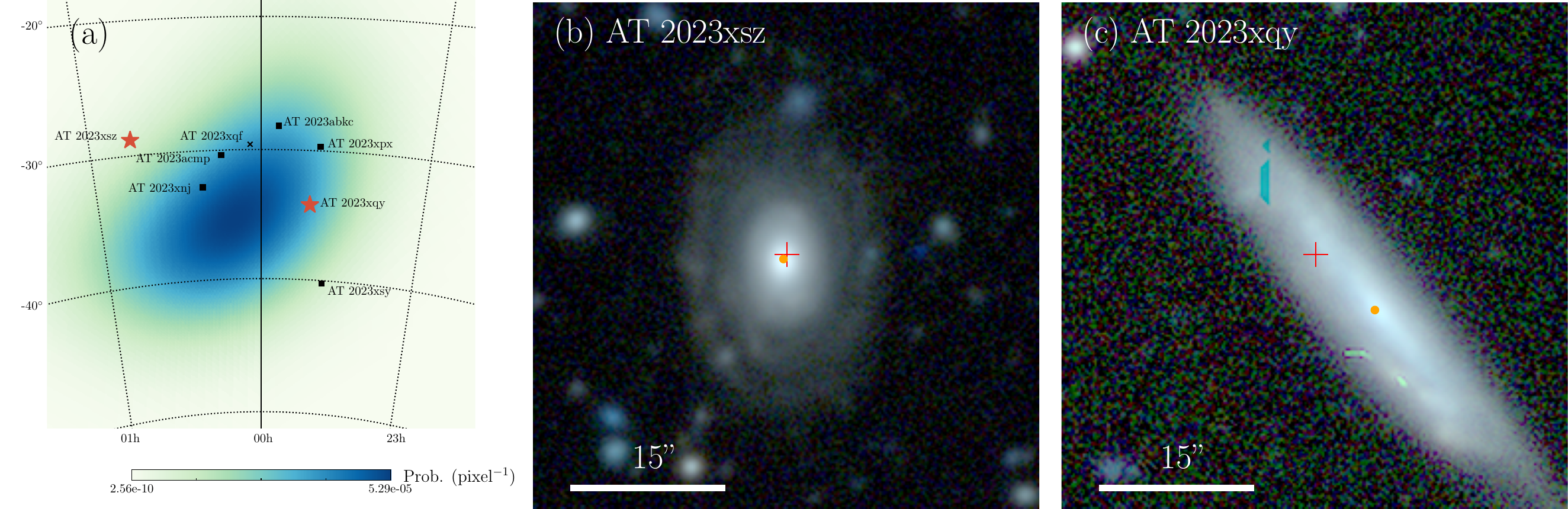}
\caption{(a) Sky locations of 8 optical candidates. The background color represents the spatial probability distribution of the gravitational-wave event derived from the GW signal, with blue shades indicating higher probability. The cross mark for AT2023xqf indicates that no host galaxy was found in the cross-match. AT2023xsz and AT2023xqy are marked with red stars, denoting that the redshifts of their host galaxies are consistent with the values reported by \cite{Niu25}. Other candidates are marked with black squares. (b) Location of the optical transient AT2023xsz within its host galaxy, PGC 737216. The transient and the host galaxy are marked with a red cross and an orange point, respectively. The background image uses DES DR10 imaging data \citep{2016MNRAS.460.1270D}. (c) Similar to (b) but for AT2023xqy within its host galaxy, PGC 72135. Additionally, two defects are visible in the DES image, with cyan speckles in the upper-left and light-green streaks in the lower-right corner; however, they do not affect our presentation of AT2023xqy and its host galaxy.
\label{fig:skymap}}
\end{figure*}

\begin{figure*}
\plotone{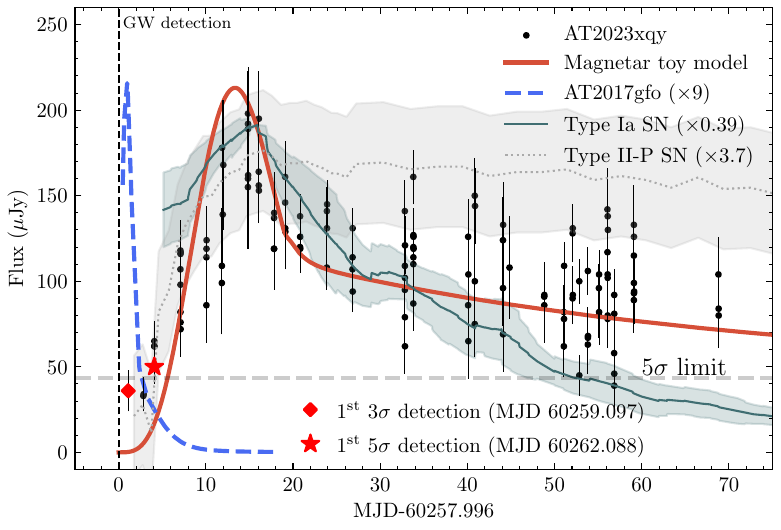}
\caption{The ATLAS \textit{o}-band light curve of AT2023xqy with a magnetar toy-model match (solid red line) is compared with a Johnson \textit{R}-band Type Ia template (cyan solid line with confidence interval), a ZTF \textit{r}-band Type IIP template (grey dashed line with confidence interval), and the theoretical \textit{r}-band kilonova light curve of AT2017gfo (blue dashed line). All photometric measurements of AT2023xqy exceed the $3\sigma$ detection threshold, with the first $3\sigma$ and $5\sigma$ detections indicated in the plot. For reference, the $5\sigma$ limiting depth under dark-time conditions is $43.65~\mu{\rm Jy}$ ($m \approx 19.8$), which is also indicated. The AT2023xqy curve shows the measured fluxes of the source, while the comparison templates have been peak-normalized to highlight differences in light-curve morphology. The Type Ia template is from \cite{2006AJ....131..527J}, the Type IIP curve from Figure~7 of \cite{das2025lowluminositytypeiipsupernovae}, and the kilonova model from \cite{Cowperthwaite_2017}.}
\label{fig:fit}
\end{figure*}

\section{Searching for Electromagnetic Counterparts to GW231109\_235456}
\label{sec:seraching}

We performed an initial search for EM counterparts using the Transient Name Server\footnote{\url{https://www.wis-tns.org/}}. The search region was defined by the all-sky spatial probability density map of GW231109\_235456 provided by GraceDB\footnote{\url{https://gracedb.ligo.org/}}, selecting the area that encloses 90\% of the probability. Since both numerical simulations and observations of the only confirmed GWEM counterpart, AT2017gfo, show that emissions from binary neutron star mergers rise steeply and peak within a few days \citep{LIGOScientific:2017vwq,Cowperthwaite_2017}, we further restricted our candidates to transients reported within four days of GW231109\_235456. Specifically, we considered reports between MJD~60257.996 (the detection time of GW231109\_235456) and MJD~60261.996.

In parallel, we examined high-energy counterparts using the Fermi-GBM catalog\footnote{\url{https://heasarc.gsfc.nasa.gov/W3Browse/fermi/fermigbrst.html}}. No burst was found during this interval, except for GRB~231109A (trigger 231109274), which occurred about 17~hours prior to GW231109\_235456. Given the temporal offset, it is unlikely to be associated.

By combining the 90\% probability sky region and excluding one confirmed Type~Ia supernova, we identified a total of eight optical candidates. Of these, four were discovered by the Gravitational-wave Optical Transient Observer \citep[GOTO, ][]{dyer2024gravitationalwaveopticaltransientobserver}, two by Asteroid Terrestrial-impact Last Alert System \citep[ATLAS, ][]{2018PASP..130f4505T}, and one by Pan-STARRS \citep{2002SPIE.4836..154K}. An additional candidate, AT2023acmp, was reported by JWST/NIRCam, though the discovery filter was not specified. Except for AT2023acmp, the discovery magnitudes of the remaining seven candidates were all in the range of 19–20~mag. The details of these candidates are listed in Table~1, and their spatial distributions are shown in Figure~\ref{fig:skymap}(a).

After cross-matching with the {\tt HyperLEDA} galaxy catalog \citep{2003A&A...412...45P} within a 5-arcminute radius, seven optical candidates, including AT2023acmp, which was already identified with its host galaxy, were found to be associated with known galaxies. All of these hosts have either DES~DR9 photometric redshift estimates or spectroscopic redshift measurements, as summarized in Table~1. \citet{Niu25} suggested that the BNS merger lies at a distance of $165^{+70}_{-69}$~Mpc, corresponding to a redshift of $0.04 \pm 0.02$. Based on this, we focused on candidates with host galaxy redshifts below 0.1. This criterion excluded five candidates with higher redshifts, leaving two transients: AT2023xsz with a host DES~DR9 photometric redshift of 0.070 \citep{2014MNRAS.445.1482S, 2016MNRAS.460.1270D}, and AT2023xqy with a host 2dF spectroscopic redshift of 0.0390 \citep{2001MNRAS.328.1039C}.

\subsection{AT2023xsz}

AT2023xsz was first reported by the ATLAS on 2023 November 12. Cross-matching with {\tt HyperLEDA} places it 0.52~arcseconds (corresponding to 0.7~kpc at its host galaxy redshift) from the center of its host galaxy, PGC~737216 [Figure~\ref{fig:skymap}(b)]. In DES images, PGC~737216 appears as a disk galaxy with a ring or spiral arms. DES~DR9 provides a photometric redshift of $z=0.070\pm0.013$, corresponding to a luminosity distance of $\sim$330~Mpc, slightly beyond the 2$\sigma$ distance estimate for GW231109\_235456 reported by \citet{Niu25}. 

The discovery magnitude reported to TNS was 19.114~mag in the ATLAS cyan ($c$) band. The ATLAS light curve \citep{2020PASP..132h5002S} shows a peak magnitude of $19.35\pm0.30$~mag in the orange ($o$) band (see Figure~\ref{fig:fit_xsz}), while the cyan band is sparsely sampled. The $o$-band light curve indicates that AT2023xsz had already brightened by MJD~60259.135 and reached peak brightness near MJD~60262, suggesting a rise time slightly longer than 3~days. However, as its brightness was close to the ATLAS 5$\sigma$ detection limit ($\sim$19~mag), the data quality is relatively poor. Without correcting for reddening, the absolute magnitude at peak is approximately $-18.24$~mag in the $o$-band, and the light curve roughly resembles that of a Type~Ia supernova, albeit with significant scatter. Considering possible reddening, the peak absolute magnitude and overall light curve shape are consistent with a Type~Ia supernova (see Figure~\ref{fig:fit_xsz}); however, the data are too noisy to draw a definitive conclusion for AT2023xsz.

\subsection{AT2023xqy}
\label{sec:at2023}
AT2023xqy was first reported by ATLAS on November 13, 2023. Cross-matching with {\tt HyperLEDA} indicated a projected offset of 9.86 arcseconds from the host galaxy's core. The host galaxy of AT2023xqy is identified as PGC 72135, also known as ESO 408-16. This edge-on disk galaxy has a core that appears blurred due to extinction, suggesting that the core position provided by {\tt HyperLEDA} may be inaccurate. Re-measuring the angular distance between AT2023xqy and the host galaxy core using DES DR10 $i$-band images yields approximately 8.60 arcseconds [Figure \ref{fig:skymap}(c)]. PGC 72135 has a spectroscopic redshift measurement from 2dF of 0.0390, corresponding to a luminosity distance of 178.6 Mpc, which is in good agreement with the 165 Mpc distance estimated by \citet{Niu25}. We find that the projected distance of AT2023xqy from the host galactic center of approximately 6.9 kpc, making a tidal disruption event unlikely, as such events typically occur in the immediate vicinity of the galactic nucleus \citep{2020SSRv..216..124V}. The above coincidences indicate that AT2023xqy may be a plausible candidate for the EM counterpart of BNS merger candidate GW231109\_235456. 

The discovery magnitude of AT2023xqy reported on TNS was 19.32~mag in the GOTO $L$ band. We also extracted its light curve from the high-cadence ATLAS data. Similar to AT2023xsz, AT2023xqy lacks observations in the $c$-band, so we adopted ATLAS $o$-band measurements. The light curve shows that AT2023xqy began brightening after its first $3\sigma$ detection at MJD 60259.097 and was securely detected above $5\sigma$ at MJD 60262.088, followed by a prolonged $\sim$15-day rise to peak brightness. The peak is relatively sharp, with the ATLAS $o$-band magnitude reaching 18.31$\pm$0.11~mag. After the peak, the luminosity declined rapidly over $\sim$5~days, followed by a plateau phase lasting at least 50~days. Due to a seasonal gap, no ATLAS observations were obtained between MJD~60327.828 and 60424.145; by the end of this interval, AT2023xqy had faded below the ATLAS detection limit.

\section{Possible origin of AT2023\MakeLowercase{xqy} }
\label{sec:scenario}

Given the high-quality observation on the shape of the AT2023xqy light curve, we examine two possible scenarios to explain the origin of this transient phenomenon, a BNS merger-irrelevant scenario involving a peculiar supernova or  a BNS merger-relevant scenario involving a magnetar-powered kilonova. 

\subsection{A BNS merger-irrelevant scenario?}

The long-lasting ($\sim100$ days) light curve evolution may naturally suggest us a BNS merger-irrelevant supernova origin of AT2023xqy. However, as shown in Figure~\ref{fig:fit}, we find that its peculiar feature may not be well-matched with typical supernova types. On the one hand, the peak absolute magnitude of AT2023xqy is $-17.95$~mag in the $o$-band, significantly fainter than that of Type~Ia supernovae, even without reddening correction. On the other hand, AT2023xqy exhibits a plateau of 50-days, which may not be seen in Type Ib/Ic/II-L supernovae which display a rapidly declining single-peaked light curve feature after maximum luminosity \citep{Taddia_2015}. Though this plateau may be observed in some superluminous supernovae (SLSNe), explained by a model with a central magnetar engine \citep[e.g.,][]{2017ApJ...850...55N}, their peak luminosity is an order of magnitude higher than that of AT2023xqy \citep{Gal_Yam_2019}, making it an unfavorable explanation. 

The closest conventional analog Type II-P supernovae  displaying comparable plateau features, also fails to fully account for the observed behavior.  In Figure~\ref{fig:fit}, we compare its light curve to Type~II-P supernova templates recently constructed from 330 Type~II-P supernovae observed by the Zwicky Transient Facility Census of the Local Universe survey \citep{2025PASP..137d4203D}. Still, we observe a difference that AT2023xqy declines by $\sim$0.6~mag from peak within $\sim$5~days, whereas Type~II-P supernovae typically exhibit only very weak declines ($\sim 0.1$~mag) over the same period, which suggests that AT2023xqy does not resemble a typical Type~II-P supernova.

Furthermore, we estimate the expected number of supernovae within the comoving volume of GW231109\_235456 probed by LVK observations between the GW trigger time (MJD 60257.996) and the onset of AT2023xqy (MJD 60259.097–60262.088). Adopting the local volumetric supernova rates  reported in \cite{2025A&A...698A.306M}, which account for both Type I and II supernovae, we find an expectation value of $\sim 0.01–0.05$ events within two to four days after the BNS merger in this volume. Under the assumption of Poissonian statistics, this corresponds to a relatively small probability of $1.0\%–4.7\%$ for a temporally coincident supernova occurring within the designated spatiotemporal window.

\subsection{A BNS merger-relevant scenario?}

As outlined in Section~\ref{sec:at2023}, the spatial alignment of AT2023xqy including its luminosity distance and rise time closely matches the localization constraints provided by the LVK GraceDB alert for GW231109\_235456. Furthermore, compact binary population synthesis models integrated into cosmological hydrodynamic simulations \citep[e.g.,][]{2022MNRAS.509.1557C} predicted a preferential association of BNS mergers with spiral galaxies of stellar mass around $10^{10} M_{\odot}$ at redshifts $z<0.1$. Intriguingly, the host galaxy of AT2023xqy, ESO 408-16 (PGC 72135), is classified as an Sc-type galaxy with an absolute magnitude of $-21.05$ mag \citep{1989spce.book.....L}, corresponding to a typical stellar mass of $\sim 1.4\times 10^{10}M_{\odot}$, which is consistent with the above prediction. Additionally, BNS mergers typically exhibit spatial offsets from their host galaxy centers due to natal kicks imparted during the second supernova. AT2023xqy is offset by $\sim 6.9$ kpc from the center of ESO 408-16, a value well-aligned with the empirical distribution \citep{2023ApJ...947...59L} derived from 27 short gamma-ray burst (sGRB) offsets reported in \citet{2010ApJ...708....9F} and \citet{2013ApJ...776...18F}. Taking together, these observational coincidences suggest that AT2023xqy may be a plausible EM counterpart candidate for the BNS merger GW231109\_235456.

In principle, the main EM counterparts of BNS mergers in the Optical band are: (1) kilonovae, powered by the radioactive decay of the heavy elements, such as lanthanide elements synthesized via the $r$-process in the neutron-rich material ejecta \citep[e.g.,][]{1998ApJ...507L..59L,2010MNRAS.406.2650M,2017Natur.551...80K}; (2) afterglows, powered by the shock between the relativistic jet launched by the central engine and their ambient interstellar medium \citep[e.g.,][]{2021ApJ...922..154M}. However, the $o$-band peak magnitude of AT2023xqy $\sim 18 ~\rm mag$ is significantly brighter than the typical luminosity for both kilonova and afterglow signals placed at the same distance. For example, if we place the kilonova and afterglow associated with GW170817 at $\rm d_{\rm L}~\sim 165~\rm Mpc$, their peak magnitude is about $\sim 20/29$ mag. Moreover, the peak time of AT2023xqy is about $\sim 15$ days after the GW alert, which is significantly longer than the typical kilonova peak time of $\sim 1-2$ days but also significantly shorter than the typical afterglow peak time of $\sim 100-200$ days. Owing to the above two reasons, we conclude that the light curve of AT2023xqy is difficult to reconcile with  standard kilonova and afterglow signals. 

Here we propose a possible magnetar-powered kilonova scenario to explain the light curve of AT2023xqy.  In principle, the remnant of the BNS merger may be a rapidly rotating magnetar or a black hole \citep[e.g., ][]{2008PhRvD..78h4033B,2017ApJ...844L..19P, 2022ApJ...939...51M, 2022A&A...666A.174S,2025chen}, depending on the component masses and equation of state (EOS) of BNS systems, which has been supported by both general relativistic hydrodynamic (GRMHD) simulations \citep[e.g.,][]{ 2017PhRvL.119w1102S,2025ApJ...984L..61M}  and observations on sGRBs, including the extended emission \citep[e.g.,][]{2008MNRAS.385.1455M}, X-ray flares \citep[e.g.,][]{2006Natur.442.1008C} and the internal plateaus \citep[e.g.,][]{2023arXiv230705689S}. In this magnetar scenario, the spin-down of the remnant magnetar supplies an additional energy through the strong magnetic wind \citep[e.g., ][]{2006Sci...311.1127D, 2006MNRAS.372L..19F, 2013ApJ...771L..26G, 2013ApJ...776L..40Y}, and thus may enhance the luminosity of the kilonova and extend its lasting time. By the toy model that the magnetar is surrounded by a quasi-spherical ejecta shell (more details can be seen in Appendix \ref{app:model}), we find that the light curve of AT2023xqy may be explained by a central magnetar with typical spin period of $P_{i}\sim 27~\rm ms$ and magnetic field of $B \sim 1\times 10^{15}\rm~G$ \citep[e.g.,][]{PhysRevD.93.044065} with a flat temperature of $\sim 3860~\rm K$, which can be seen in the red solid line in Figure~\ref{fig:fit}. The settings of other parameters including the mass $m_{\rm ej}$ and velocity $v_{\rm ej}$ of the ejecta can be seen in Table~\ref{table:para}. The first 15-day rise of the light curve comes from the blackbody radiation of the expanding ejecta, heated by both the radioactive decay and the magnetic wind from the magnetar and the subsequent rapid decay of 5-days is attributed by the ejecta cools down as it expands till reaching the floor temperature. The plateau emission lasting for at least 50 days is mainly powered by the stable spin-down energy injection. 

Nevertheless, we should point out that there is an nonnegligible caveat for the magnetar-powered kilonova scenario on the lifetime of the magnetar. Currently, the observations on sGRBs only provide the evidence on the existence of magnetar merger remnant with $\tau_{\rm life}$ of several hundreds of seconds \citep[e.g.,][]{2015ApJ...805...89L}. However, the lifetime $\tau_{\rm life}$ needed to recover the light curve of AT2023xqy should be within the range of $\sim [70,160]$ days, otherwise the late plateau should suffer a sharp decay quickly or remains bright for ATLAS detection. One possible explanation is that the magnetar is a long-lived marginally-stable supramassive NS with mass $1.0M_{\rm TOV}<M_{\rm NS}\lesssim 1.2M_{\rm TOV}$ \citep[e.g.,][]{2016MNRAS.459..646B}. By the procedure given in \citet{2022A&A...666A.174S}, we find that with the total mass of $\rm GW231109\_235456$ reported in \citet{Niu25} $M_{\rm tot}\sim 2.9M_{\odot}$, the mass of the merger remnant $M_{\rm NS}$ is around $\sim 2.6-2.8M_{\odot}$ depending on the mass ratio $q$, indicating a EOS of NS with large $M_{\rm TOV}$ such as AB-N and AB-L \citep[e.g.,][]{1977ApJS...33..415A}, which is however much stiffer than that constrained by current available multi-messenger observations \citep[e.g.,][]{2020Sci...370.1450D}. 

While the observed light curve can be reproduced by our model, we propose AT2023xqy as a plausible candidate for a magnetar-powered kilonova under extreme conditions, noting that this interpretation requires further observational confirmation. Specifically a further follow-up observation in the radio band will be significantly helpful in determining the nature of AT2023xqy, since the radio afterglow signals may also been enhanced by the existence of the magnetar central engine i.e., several ten times brighter than the standard afterglow \citep{2015ApJ...807..163G}, thus providing a critical diagnostics to either support or challenge this interpretation.

\section{Summary} 
\label{sec:summary}

In this letter, we carried out a systematic archival search for the EM counterparts of the BNS merger candidate GW231109\_235456. By examining all transients reported within four days of the event and inside the 90\% sky localization, we identified eight candidates. Among them, two (AT2023xsz and AT2023xqy) have host galaxies consistent with the inferred merger distance. 

AT2023xqy, located at 178.6~Mpc, shows the best spatial and distance agreement with GW231109\_235456. Furthermore, near the trigger time of GW231109\_235456 (MJD~60257.996), AT2023xqy showed evidence of a $\sim$15-day rise, first detected at $3\sigma$ significance on MJD~60259.097 and confirmed above $5\sigma$ on MJD~60262.088. This was followed by a rapid $\sim$5-day decline and a plateau lasting at least 50~days, with the subsequent decay unobserved due to a data gap. The spatiotemporal coincidences indicate that AT2023xqy could be a candidate for the EM counterpart of BNS merger candidate GW231109\_235456, though its lightcurve is difficult to reconcile with a standard kilonova.

We examine two possible scenarios for the origin of AT2023xqy, a BNS merger-irrelevant scenario involving a supernova or a BNS merger-relevant scenario involving a magnetar-powered kilonova. In the former scenario, AT2023xqy may not be clearly categorized to a certain type of known supernovae, though with close analog of Type II-P. In the latter scenario, a simple toy model suggests that the emission could arise from ejecta heated by both radioactive decay and the spin-down energy of a nascent magnetar with an initial spin period of $\sim27$\,ms and a magnetic field of $\sim$$10^{15}$~G. 

With strong coincidence in both occurrence time and spatial distribution, though a peculiar supernova origin cannot be ruled out, AT2023xqy remains a plausible EM counterpart candidate to GW231109\_235456, potentially representing the second GW EM counterpart candidate following AT2017gfo. This case, if confirmed, may indicate that the EM counterparts of BNS merger GW events exhibit broad diversity compared with GW170817, which may be considered in future searches. Follow-up radio observations are strongly encouraged, as they may provide critical diagnostics for the magnetar-powered interpretation.

\section*{acknowledgments}
YH acknowledges the support from the National Science Foundation of China (NSFC grant No. 12422303), the Fundamental Research Funds for the Central Universities (grant Nos. 118900M122, E5EQ3301X2, and E4EQ3301X2), and the National Key R\&D Programme of China (grant No. 2023YFA1608303). ZC acknowledges the Postdoctoral Fellowship Program of CPSF under Grant Number GZB20250735. YL acknowledges the support from the Strategic Priority Program of the Chinese Academy of Sciences (grant no. XDB 23040100), the National Natural Science Foundation of China (grant nos. 12273050 and 12533009), and the National Astronomical Observatory of China (grant no. E4TG660101). JFL acknowledges the support from the NSFC through grant Nos. of 11988101 and 11933004, and the New Cornerstone Science Foundation through the New Cornerstone Investigator Program and the XPLORER PRIZE.

This research has made use of the CfA Supernova Archive, which is funded in part by the National Science Foundation through grant AST 0907903.

\appendix
\setcounter{table}{0}   
\setcounter{figure}{0}
\renewcommand{\thetable}{A\arabic{table}}
\renewcommand{\thefigure}{A\arabic{figure}}
\section{Toy model of Magnetar-powered Kilonova}
\label{app:model}

Suppose the magnetar is surrounded by a quasi-spherical ejecta shell with mass $M_{\rm ej}$, the total kinetic energy of the system can be expressed as \citep[e.g.,][]{2013ApJ...776L..40Y,2015ApJ...807..163G}
\begin{equation}
E=(\Gamma-1)M_{\rm ej}+\Gamma E'_{\rm int}+(\Gamma^2-1)M_{\rm sw},
\end{equation}
where $\Gamma$ is the bulk Lorentz factor of the ejecta, $M_{\rm sw}$ is the swept mass from the interstellar medium, $R$ is the radius of the ejecta, and $E_{\rm int}$ is the internal energy. Here and hereafter the symbols with prime represent the quantities measured in the comoving rest frame. Then the dynamical evolution of the ejecta can be determined with the following ordinary differential equations (ODEs)
\begin{eqnarray}
{d\Gamma\over dt}={{dE\over dt}-\Gamma {\cal D}\left({dE'_{\rm int}\over
dt'}\right)-(\Gamma^2-1)c^2\left({dM_{\rm sw}\over dt}\right)\over
M_{\rm ej}c^2+E'_{\rm int}+2\Gamma M_{\rm sw}c^2},
\label{eq:Gt}
\end{eqnarray}
\begin{eqnarray}
{dE\over dt}= L_{\rm sd}+{\cal D}^{2}L'_{\rm ra}-{\cal D}^{2}L'_{\rm e}.
\label{eq:Et}
\end{eqnarray}
\begin{eqnarray}
{dE'_{\rm int}\over dt'}= {\cal D}^{-2}L_{\rm sd}+ L'_{\rm ra} -L'_{\rm e}
-\frac{E'_{\rm int}}{3V'}{dV'\over dt'},
\label{eq:Ep}
\end{eqnarray}
\begin{eqnarray}
{dV'\over dt'}=4\pi R^2\beta c,
\label{eq:Vp}
\end{eqnarray}
\begin{eqnarray}
{dR\over dt}={\beta c\over (1-\beta)}.
\label{eq:rt}
\end{eqnarray}
In the above equations, ${\cal D}=1/[\Gamma(1-\beta)]$ is the Doppler factor with $\beta=\sqrt{1-\Gamma^{-2}}$, and the comoving time $dt'$ can be estimated from the observer frame time $dt$ by $dt'={\cal D}dt$. 
The term $L_{\rm ra}$ is the r-process power given by
\begin{eqnarray}
L'_{\rm ra}=4\times10^{49}M_{\rm ej,-2}\left[{1\over2}-{1\over\pi}\arctan \left({t'-t'_0\over
t'_\sigma}\right)\right]^{1.3}~\rm erg~s^{-1},
\label{eq:Lrap}
\end{eqnarray}
with $t'_0 \sim 1.3$\,s and $t'_\sigma \sim 0.11$\,s \citep[e.g.,][]{2012MNRAS.426.1940K} and hereafter the convention $Q_x=Q/10^x$ is adopted in cgs units.  The term $L'_e$ is the luminosity emitted, which can be written as
\begin{eqnarray}
L'_e=\left\{
\begin{array}{l l}
  {E'_{\rm int}c\over \tau R/\Gamma}, & \tau>1, \\
  {E'_{\rm int}c\over R/\Gamma}, &\tau<1,\\ \end{array} \right.\
  \label{eq:Lep}
\end{eqnarray}
where $\tau=\kappa (M_{\rm ej}/V')(R/\Gamma)$ is the optical depth of the ejecta with opacity $\kappa$ \citep[e.g., ][]{2010ApJ...717..245K}. The term $L_{\rm sd}$ is the spin-down energy of the central magnetar remnant assuming an isotropic Poynting flux, which can be written as
\begin{eqnarray}
L_{\rm sd}=L_{\rm sd,i}\left(1+{t\over t_{\rm
sd}}\right)^{-2},
\end{eqnarray}
where
%
%\begin{eqnarray}
$L_{\rm sd,i}=10^{47}~R_{s,6}^6B_{14}^{2}P_{i,-3}^{-4}\rm~erg~s^{-1}$
%\end{eqnarray}
is the initial spin-down luminosity, and
%\begin{eqnarray}
$t_{\rm sd}=2\times10^{5}~R_{s,6}^{-6}B_{14}^{-2}P_{i,-3}^{2}~\rm s$
%\end{eqnarray}
is the spin-down timescale \citep[e.g., ][]{2006Sci...311.1127D, 2006MNRAS.372L..19F, 2013ApJ...771L..26G, 2013ApJ...776L..40Y}, with $R_{s}$, $B$ and $P_{i}$ representing the radius, magnetic field, and initial spin period of the magnetar. 
Then, the temperature of the expanding ejecta can be estimated directly by $T'=\big({E'_{\rm int}}/{aV'}\big)^{1/{4}}$, where $a$ is the radiation constant. Assuming a blackbody spectrum for the
thermal emission of the mergernova, for a certain observational frequency $\nu$, the observed flux can be calculated as
\begin{eqnarray}
F_{\nu}={1\over4\pi D_L^2\max(\tau,1)}{8\pi^2  {\cal D}^2R^2\over
h^3c^2\nu}{(h\nu/{\cal D})^4\over \exp(h\nu/{\cal D}kT')-1},
\end{eqnarray}
where $h$ is the Planck constant and  $D_{\rm L}$ is the luminosity distance.  Following the supernova modeling with magnetar central engine \citep[e.g.,][]{2017ApJ...850...55N}, we introduce a flat temperature $T_{f}$, i.e.,  once the temperature $T'$ decreases to $T_{f}$ with the expansion of the ejecta, the temperature will remain constant (i.e., isothermal expansion) and the ejecta radius evolves with $R=(3E_{\rm int}/4\pi a T_{f}^4)^{1/3} $ after this value.

\begin{table*}
\caption{Model parameters adopted for recovering the light curves of AT2023xqy}
\label{table:para} 
\centering
\begin{tabular}{lcc}    \hline
        Parameter & Name & Value \\
        \hline
       $m_{\rm ej}$ & Ejecta mass & $4\times 10^{-3} M_{\odot}$  \\ 
       $v_{\rm ej}$ & Ejecta velocity & $0.1~c$ \\
       $\kappa$ & Opacity & $10 \rm ~g/cm^2$ \\
       $P_{i}$ & Magnetar period & $\rm 27~ms$ \\
       
       $B$ & Magnetic field & $\rm 10^{15}~G$ \\
       
       $R_s$ & Magnetar radii & $11~\rm km$ \\
        $T_f$ & Floor Temperature & $3860~\rm K$ \\
       \hline    
\end{tabular}
\end{table*}

\section{Light curve of AT2023xsz}
\begin{figure*}
\plotone{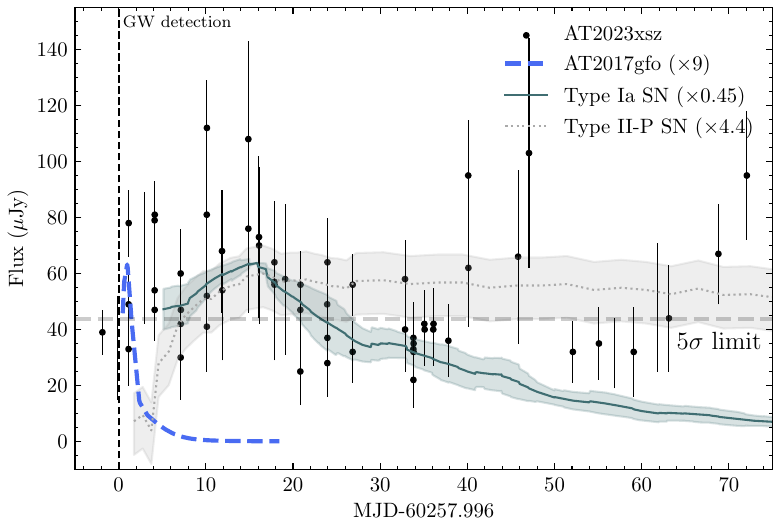}
\caption{Similar to Figure \ref{fig:fit} but for AT2023xsz. However, we did not apply our toy model to AT2023xsz due to its relatively low photometric quality. Only photometric measurements above the $2\sigma$ detection threshold are shown here.}
\label{fig:fit_xsz}
\end{figure*}

Figure~\ref{fig:fit_xsz} shows the light curve of AT2023xsz, compared with template light curves of various transient types.

\bibliography{references}

\begin{thebibliography}{}
\expandafter\ifx\csname natexlab\endcsname\relax\def\natexlab#1{#1}\fi
\providecommand{\url}[1]{\href{#1}{#1}}
\providecommand{\dodoi}[1]{doi:~\href{http://doi.org/#1}{\nolinkurl{#1}}}
\providecommand{\doeprint}[1]{\href{http://ascl.net/#1}{\nolinkurl{http://ascl.net/#1}}}
\providecommand{\doarXiv}[1]{\href{https://arxiv.org/abs/#1}{\nolinkurl{https://arxiv.org/abs/#1}}}

\bibitem[{Aasi {et~al.}(2015)}]{LIGOScientific:2014pky}
Aasi, J., {et~al.} 2015, Class. Quant. Grav., 32, 074001,
  \dodoi{10.1088/0264-9381/32/7/074001}

\bibitem[{Abac {et~al.}(2025)}]{gwtc4}
Abac, A.~G., {et~al.} 2025.
\newblock \doarXiv{2508.18082}

\bibitem[{Abbott {et~al.}(2017{\natexlab{a}})}]{LIGOScientific:2017vwq}
Abbott, B.~P., {et~al.} 2017{\natexlab{a}}, Phys. Rev. Lett., 119, 161101,
  \dodoi{10.1103/PhysRevLett.119.161101}

\bibitem[{Abbott {et~al.}(2017{\natexlab{b}})}]{LIGOScientific:2017ync}
---. 2017{\natexlab{b}}, Astrophys. J. Lett., 848, L12,
  \dodoi{10.3847/2041-8213/aa91c9}

\bibitem[{Abbott {et~al.}(2017{\natexlab{c}})Abbott, {Abbott}, {Abbott},
  {Acernese}, {Ackley}, {Adams}, {Adams}, {Addesso}, {Adhikari}, {Adya},
  {Affeldt}, {Afrough}, {Agarwal}, {Agathos}, {Agatsuma}, {Aggarwal}, {Aguiar},
  {Aiello}, {Ain}, {Ajith}, {Allen}, {Allen}, {Allocca}, {Altin}, {Amato},
  {Ananyeva}, {Anderson}, {Anderson}, {Angelova}, {Antier}, {Appert}, {Arai},
  {Araya}, {Areeda}, {Arnaud}, {Arun}, {Ascenzi}, {Ashton}, {Ast}, {Aston},
  {Astone}, {Atallah}, {Aufmuth}, {Aulbert}, {AultONeal}, {Austin},
  {Avila-Alvarez}, {Babak}, {Bacon}, {Bader}, {Bae}, {Baker}, {Baldaccini},
  {Ballardin}, {Ballmer}, {Banagiri}, {Barayoga}, {Barclay}, {Barish},
  {Barker}, {Barkett}, {Barone}, {Barr}, {Barsotti}, {Barsuglia}, {Barta},
  {Barthelmy}, {Bartlett}, {Bartos}, {Bassiri}, {Basti}, {Batch}, {Bawaj},
  {Bayley}, {Bazzan}, {B{\'e}csy}, {Beer}, {Bejger}, {Belahcene}, {Bell},
  {Berger}, {Bergmann}, {Bero}, {Berry}, {Bersanetti}, {Bertolini},
  {Betzwieser}, {Bhagwat}, {Bhandare}, {Bilenko}, {Billingsley}, {Billman},
  {Birch}, {Birney}, {Birnholtz}, {Biscans}, {Biscoveanu}, {Bisht}, {Bitossi},
  {Biwer}, {Bizouard}, {Blackburn}, {Blackman}, {Blair}, {Blair}, {Blair},
  {Bloemen}, {Bock}, {Bode}, {Boer}, {Bogaert}, {Bohe}, {Bondu}, {Bonilla},
  {Bonnand}, {Boom}, {Bork}, {Boschi}, {Bose}, {Bossie}, {Bouffanais}, {Bozzi},
  {Bradaschia}, {Brady}, {Branchesi}, {Brau}, {Briant}, {Brillet}, {Brinkmann},
  {Brisson}, {Brockill}, {Broida}, {Brooks}, {Brown}, {Brown}, {Brunett},
  {Buchanan}, {Buikema}, {Bulik}, {Bulten}, {Buonanno}, {Buskulic}, {Buy},
  {Byer}, {Cabero}, {Cadonati}, {Cagnoli}, {Cahillane}, {Calder{\'o}n
  Bustillo}, {Callister}, {Calloni}, {Camp}, {Canepa}, {Canizares}, {Cannon},
  {Cao}, {Cao}, {Capano}, {Capocasa}, {Carbognani}, {Caride}, {Carney},
  {Casanueva Diaz}, {Casentini}, {Caudill}, {Cavagli{\`a}}, {Cavalier},
  {Cavalieri}, {Cella}, {Cepeda}, {Cerd{\'a}-Dur{\'a}n}, {Cerretani},
  {Cesarini}, {Chamberlin}, {Chan}, {Chao}, {Charlton}, {Chase},
  {Chassande-Mottin}, {Chatterjee}, {Chatziioannou}, {Cheeseboro}, {Chen},
  {Chen}, {Chen}, {Cheng}, {Chia}, {Chincarini}, {Chiummo}, {Chmiel}, {Cho},
  {Cho}, {Chow}, {Christensen}, {Chu}, {Chua}, {Chua}, {Chung}, {Chung}, \&
  {Ciani}}]{2017ApJ...848L..12A}
Abbott, B.~P., {Abbott}, R., {Abbott}, T.~D., {et~al.} 2017{\natexlab{c}},
  \apjl, 848, L12, \dodoi{10.3847/2041-8213/aa91c9}

\bibitem[{Abbott {et~al.}(2020)}]{LIGOScientific:2020aai}
Abbott, B.~P., {et~al.} 2020, Astrophys. J. Lett., 892, L3,
  \dodoi{10.3847/2041-8213/ab75f5}

\bibitem[{Acernese {et~al.}(2015)}]{VIRGO:2014yos}
Acernese, F., {et~al.} 2015, Class. Quant. Grav., 32, 024001,
  \dodoi{10.1088/0264-9381/32/2/024001}

\bibitem[{{Arnett} \& {Bowers}(1977)}]{1977ApJS...33..415A}
{Arnett}, W.~D., \& {Bowers}, R.~L. 1977, \apjs, 33, 415,
  \dodoi{10.1086/190434}

\bibitem[{{Baiotti} {et~al.}(2008){Baiotti}, {Giacomazzo}, \&
  {Rezzolla}}]{2008PhRvD..78h4033B}
{Baiotti}, L., {Giacomazzo}, B., \& {Rezzolla}, L. 2008, \prd, 78, 084033,
  \dodoi{10.1103/PhysRevD.78.084033}

\bibitem[{{Breu} \& {Rezzolla}(2016)}]{2016MNRAS.459..646B}
{Breu}, C., \& {Rezzolla}, L. 2016, \mnras, 459, 646,
  \dodoi{10.1093/mnras/stw575}

\bibitem[{{Campana} {et~al.}(2006){Campana}, {Mangano}, {Blustin}, {Brown},
  {Burrows}, {Chincarini}, {Cummings}, {Cusumano}, {Della Valle}, {Malesani},
  {M{\'e}sz{\'a}ros}, {Nousek}, {Page}, {Sakamoto}, {Waxman}, {Zhang}, {Dai},
  {Gehrels}, {Immler}, {Marshall}, {Mason}, {Moretti}, {O'Brien}, {Osborne},
  {Page}, {Romano}, {Roming}, {Tagliaferri}, {Cominsky}, {Giommi}, {Godet},
  {Kennea}, {Krimm}, {Angelini}, {Barthelmy}, {Boyd}, {Palmer}, {Wells}, \&
  {White}}]{2006Natur.442.1008C}
{Campana}, S., {Mangano}, V., {Blustin}, A.~J., {et~al.} 2006, \nat, 442, 1008,
  \dodoi{10.1038/nature04892}

\bibitem[{Capote {et~al.}(2025)}]{Capote:2024rmo}
Capote, E., {et~al.} 2025, Phys. Rev. D, 111, 062002,
  \dodoi{10.1103/PhysRevD.111.062002}

\bibitem[{{Chen} {et~al.}(2025){Chen}, {Lu}, {Ma}, \& {Chu}}]{2025chen}
{Chen}, Z., {Lu}, Y., {Ma}, H., \& {Chu}, Q. 2025, submitted to MNRAS

\bibitem[{{Chu} {et~al.}(2022){Chu}, {Yu}, \& {Lu}}]{2022MNRAS.509.1557C}
{Chu}, Q., {Yu}, S., \& {Lu}, Y. 2022, \mnras, 509, 1557,
  \dodoi{10.1093/mnras/stab2882}

\bibitem[{{Colless} {et~al.}(2001){Colless}, {Dalton}, {Maddox}, {Sutherland},
  {Norberg}, {Cole}, {Bland-Hawthorn}, {Bridges}, {Cannon}, {Collins}, {Couch},
  {Cross}, {Deeley}, {De Propris}, {Driver}, {Efstathiou}, {Ellis}, {Frenk},
  {Glazebrook}, {Jackson}, {Lahav}, {Lewis}, {Lumsden}, {Madgwick}, {Peacock},
  {Peterson}, {Price}, {Seaborne}, \& {Taylor}}]{2001MNRAS.328.1039C}
{Colless}, M., {Dalton}, G., {Maddox}, S., {et~al.} 2001, \mnras, 328, 1039,
  \dodoi{10.1046/j.1365-8711.2001.04902.x}

\bibitem[{Cowperthwaite {et~al.}(2017)Cowperthwaite, Berger, Villar, Metzger,
  Nicholl, Chornock, Blanchard, Fong, Margutti, Soares-Santos, Alexander,
  Allam, Annis, Brout, Brown, Butler, Chen, Diehl, Doctor, Drout, Eftekhari,
  Farr, Finley, Foley, Frieman, Fryer, García-Bellido, Gill, Guillochon,
  Herner, Holz, Kasen, Kessler, Marriner, Matheson, Neilsen, Quataert, Palmese,
  Rest, Sako, Scolnic, Smith, Tucker, Williams, Balbinot, Carlin, Cook, Durret,
  Li, Lopes, Lourenço, Marshall, Medina, Muir, Muñoz, Sauseda, Schlegel,
  Secco, Vivas, Wester, Zenteno, Zhang, Abbott, Banerji, Bechtol, Benoit-Lévy,
  Bertin, Buckley-Geer, Burke, Capozzi, Carnero~Rosell, Carrasco~Kind,
  Castander, Crocce, Cunha, D’Andrea, Costa, Davis, DePoy, Desai, Dietrich,
  Drlica-Wagner, Eifler, Evrard, Fernandez, Flaugher, Fosalba, Gaztanaga,
  Gerdes, Giannantonio, Goldstein, Gruen, Gruendl, Gutierrez, Honscheid, Jain,
  James, Jeltema, Johnson, Johnson, Kent, Krause, Kron, Kuehn, Nuropatkin,
  Lahav, Lima, Lin, Maia, March, Martini, McMahon, Menanteau, Miller, Miquel,
  Mohr, Neilsen, Nichol, Ogando, Plazas, Roe, Romer, Roodman, Rykoff, Sanchez,
  Scarpine, Schindler, Schubnell, Sevilla-Noarbe, Smith, Smith, Sobreira,
  Suchyta, Swanson, Tarle, Thomas, Thomas, Troxel, Vikram, Walker, Wechsler,
  Weller, Yanny, \& Zuntz}]{Cowperthwaite_2017}
Cowperthwaite, P.~S., Berger, E., Villar, V.~A., {et~al.} 2017, The
  Astrophysical Journal, 848, L17, \dodoi{10.3847/2041-8213/aa8fc7}

\bibitem[{{Dai} {et~al.}(2006){Dai}, {Wang}, {Wu}, \&
  {Zhang}}]{2006Sci...311.1127D}
{Dai}, Z.~G., {Wang}, X.~Y., {Wu}, X.~F., \& {Zhang}, B. 2006, Science, 311,
  1127, \dodoi{10.1126/science.1123606}

\bibitem[{{Dark Energy Survey Collaboration} {et~al.}(2016){Dark Energy Survey
  Collaboration}, {Abbott}, {Abdalla}, {Aleksi{\'c}}, {Allam}, {Amara},
  {Bacon}, {Balbinot}, {Banerji}, {Bechtol}, {Benoit-L{\'e}vy}, {Bernstein},
  {Bertin}, {Blazek}, {Bonnett}, {Bridle}, {Brooks}, {Brunner}, {Buckley-Geer},
  {Burke}, {Caminha}, {Capozzi}, {Carlsen}, {Carnero-Rosell}, {Carollo},
  {Carrasco-Kind}, {Carretero}, {Castander}, {Clerkin}, {Collett}, {Conselice},
  {Crocce}, {Cunha}, {D'Andrea}, {da Costa}, {Davis}, {Desai}, {Diehl},
  {Dietrich}, {Dodelson}, {Doel}, {Drlica-Wagner}, {Estrada}, {Etherington},
  {Evrard}, {Fabbri}, {Finley}, {Flaugher}, {Foley}, {Fosalba}, {Frieman},
  {Garc{\'\i}a-Bellido}, {Gaztanaga}, {Gerdes}, {Giannantonio}, {Goldstein},
  {Gruen}, {Gruendl}, {Guarnieri}, {Gutierrez}, {Hartley}, {Honscheid}, {Jain},
  {James}, {Jeltema}, {Jouvel}, {Kessler}, {King}, {Kirk}, {Kron}, {Kuehn},
  {Kuropatkin}, {Lahav}, {Li}, {Lima}, {Lin}, {Maia}, {Makler}, {Manera},
  {Maraston}, {Marshall}, {Martini}, {McMahon}, {Melchior}, {Merson}, {Miller},
  {Miquel}, {Mohr}, {Morice-Atkinson}, {Naidoo}, {Neilsen}, {Nichol}, {Nord},
  {Ogando}, {Ostrovski}, {Palmese}, {Papadopoulos}, {Peiris}, {Peoples},
  {Percival}, {Plazas}, {Reed}, {Refregier}, {Romer}, {Roodman}, {Ross},
  {Rozo}, {Rykoff}, {Sadeh}, {Sako}, {S{\'a}nchez}, {Sanchez}, {Santiago},
  {Scarpine}, {Schubnell}, {Sevilla-Noarbe}, {Sheldon}, {Smith}, {Smith},
  {Soares-Santos}, {Sobreira}, {Soumagnac}, {Suchyta}, {Sullivan}, {Swanson},
  {Tarle}, {Thaler}, {Thomas}, {Thomas}, {Tucker}, {Vieira}, {Vikram},
  {Walker}, {Wechsler}, {Weller}, {Wester}, {Whiteway}, {Wilcox}, {Yanny},
  {Zhang}, \& {Zuntz}}]{2016MNRAS.460.1270D}
{Dark Energy Survey Collaboration}, {Abbott}, T., {Abdalla}, F.~B., {et~al.}
  2016, \mnras, 460, 1270, \dodoi{10.1093/mnras/stw641}

\bibitem[{Das {et~al.}(2025)Das, Kasliwal, Fremling, Sollerman, Perley, De,
  Tzanidakis, Sit, Adams, Anand, Ahumuda, Andreoni, Brennan, Brink, Bruch,
  Chen, Chu, Cook, Covarrubias, Dahiwale, Earley, Ho, Gal-Yam, Gangopadhyay,
  Hammerstein, Hinds, Karambelkar, Kong, Kulkarni, du~Laz, Liu, Meynardie,
  Miller, Nir, Patra, Pessi, Rich, Rehemtulla, Rose, Rusholme, Schulze, Sharma,
  Singh, Smith, Stein, Mandigo-Stoba, Strotjohann, Qin, Wise, Wold, Yan, Yang,
  Yao, \& Zimmerman}]{das2025lowluminositytypeiipsupernovae}
Das, K.~K., Kasliwal, M.~M., Fremling, C., {et~al.} 2025, Low-Luminosity Type
  IIP Supernovae from the Zwicky Transient Facility Census of the Local
  Universe. I: Luminosity Function, Volumetric Rate.
\newblock \doarXiv{2502.19493}

\bibitem[{{Das} {et~al.}(2025){Das}, {Kasliwal}, {Fremling}, {Sollerman},
  {Perley}, {De}, {Tzanidakis}, {Sit}, {Adams}, {Anand}, {Ahumuda}, {Andreoni},
  {Brennan}, {Brink}, {Bruch}, {Chen}, {Chu}, {Cook}, {Covarrubias},
  {Dahiwale}, {Earley}, {Ho}, {Gal-Yam}, {Gangopadhyay}, {Hammerstein},
  {Hinds}, {Karambelkar}, {Kong}, {Kulkarni}, {Jegou du Laz}, {Liu},
  {Meynardie}, {Miller}, {Nir}, {Patra}, {Pessi}, {Rich}, {Rehemtulla}, {Rose},
  {Rusholme}, {Schulze}, {Sharma}, {Singh}, {Smith}, {Stein}, {Mandigo-Stoba},
  {Strotjohann}, {Qin}, {Wise}, {Wold}, {Yan}, {Yang}, {Yao}, \&
  {Zimmerman}}]{2025PASP..137d4203D}
{Das}, K.~K., {Kasliwal}, M.~M., {Fremling}, C., {et~al.} 2025, \pasp, 137,
  044203, \dodoi{10.1088/1538-3873/adcaeb}

\bibitem[{{Dietrich} {et~al.}(2020){Dietrich}, {Coughlin}, {Pang}, {Bulla},
  {Heinzel}, {Issa}, {Tews}, \& {Antier}}]{2020Sci...370.1450D}
{Dietrich}, T., {Coughlin}, M.~W., {Pang}, P. T.~H., {et~al.} 2020, Science,
  370, 1450, \dodoi{10.1126/science.abb4317}

\bibitem[{Dyer {et~al.}(2024)Dyer, Ackley, Jiménez-Ibarra, Lyman, Ulaczyk,
  Steeghs, Galloway, Dhillon, O'Brien, Ramsay, Noysena, Kotak, Breton, Nuttall,
  Pallé, Pollacco, Killestein, Kumar, O'Neill, Kelsey, Godson, \&
  Jarvis}]{dyer2024gravitationalwaveopticaltransientobserver}
Dyer, M.~J., Ackley, K., Jiménez-Ibarra, F., {et~al.} 2024, The
  Gravitational-wave Optical Transient Observer (GOTO).
\newblock \doarXiv{2407.17176}

\bibitem[{{Fan} \& {Xu}(2006)}]{2006MNRAS.372L..19F}
{Fan}, Y.-Z., \& {Xu}, D. 2006, \mnras, 372, L19,
  \dodoi{10.1111/j.1745-3933.2006.00217.x}

\bibitem[{{Fong} \& {Berger}(2013)}]{2013ApJ...776...18F}
{Fong}, W., \& {Berger}, E. 2013, \apj, 776, 18,
  \dodoi{10.1088/0004-637X/776/1/18}

\bibitem[{{Fong} {et~al.}(2010){Fong}, {Berger}, \&
  {Fox}}]{2010ApJ...708....9F}
{Fong}, W., {Berger}, E., \& {Fox}, D.~B. 2010, \apj, 708, 9,
  \dodoi{10.1088/0004-637X/708/1/9}

\bibitem[{Gal-Yam(2019)}]{Gal_Yam_2019}
Gal-Yam, A. 2019, Annual Review of Astronomy and Astrophysics, 57, 305–333,
  \dodoi{10.1146/annurev-astro-081817-051819}

\bibitem[{Ganapathy {et~al.}(2023)}]{LIGOO4Detector:2023wmz}
Ganapathy, D., {et~al.} 2023, Phys. Rev. X, 13, 041021,
  \dodoi{10.1103/PhysRevX.13.041021}

\bibitem[{{Gao} {et~al.}(2015){Gao}, {Ding}, {Wu}, {Dai}, \&
  {Zhang}}]{2015ApJ...807..163G}
{Gao}, H., {Ding}, X., {Wu}, X.-F., {Dai}, Z.-G., \& {Zhang}, B. 2015, \apj,
  807, 163, \dodoi{10.1088/0004-637X/807/2/163}

\bibitem[{Gao {et~al.}(2016)Gao, Zhang, \& L\"u}]{PhysRevD.93.044065}
Gao, H., Zhang, B., \& L\"u, H.-J. 2016, Phys. Rev. D, 93, 044065,
  \dodoi{10.1103/PhysRevD.93.044065}

\bibitem[{{Giacomazzo} \& {Perna}(2013)}]{2013ApJ...771L..26G}
{Giacomazzo}, B., \& {Perna}, R. 2013, \apjl, 771, L26,
  \dodoi{10.1088/2041-8205/771/2/L26}

\bibitem[{{Gill} {et~al.}(2019){Gill}, {Nathanail}, \&
  {Rezzolla}}]{2019ApJ...876..139G}
{Gill}, R., {Nathanail}, A., \& {Rezzolla}, L. 2019, \apj, 876, 139,
  \dodoi{10.3847/1538-4357/ab16da}

\bibitem[{{Jha} {et~al.}(2006){Jha}, {Kirshner}, {Challis}, {Garnavich},
  {Matheson}, {Soderberg}, {Graves}, {Hicken}, {Alves}, {Arce}, {Balog},
  {Barmby}, {Barton}, {Berlind}, {Bragg}, {Brice{\~n}o}, {Brown}, {Buckley},
  {Caldwell}, {Calkins}, {Carter}, {Concannon}, {Donnelly}, {Eriksen},
  {Fabricant}, {Falco}, {Fiore}, {Garcia}, {G{\'o}mez}, {Grogin}, {Groner},
  {Groot}, {Haisch}, {Hartmann}, {Hergenrother}, {Holman}, {Huchra},
  {Jayawardhana}, {Jerius}, {Kannappan}, {Kim}, {Kleyna}, {Kochanek},
  {Koranyi}, {Krockenberger}, {Lada}, {Luhman}, {Luu}, {Macri}, {Mader},
  {Mahdavi}, {Marengo}, {Marsden}, {McLeod}, {McNamara}, {Megeath}, {Moraru},
  {Mossman}, {Muench}, {Mu{\~n}oz}, {Muzerolle}, {Naranjo}, {Nelson-Patel},
  {Pahre}, {Patten}, {Peters}, {Peters}, {Raymond}, {Rines}, {Schild},
  {Sobczak}, {Spahr}, {Stauffer}, {Stefanik}, {Szentgyorgyi}, {Tollestrup},
  {V{\"a}is{\"a}nen}, {Vikhlinin}, {Wang}, {Willner}, {Wolk}, {Zajac}, {Zhao},
  \& {Stanek}}]{2006AJ....131..527J}
{Jha}, S., {Kirshner}, R.~P., {Challis}, P., {et~al.} 2006, \aj, 131, 527,
  \dodoi{10.1086/497989}

\bibitem[{{Kaiser} {et~al.}(2002){Kaiser}, {Aussel}, {Burke}, {Boesgaard},
  {Chambers}, {Chun}, {Heasley}, {Hodapp}, {Hunt}, {Jedicke}, {Jewitt},
  {Kudritzki}, {Luppino}, {Maberry}, {Magnier}, {Monet}, {Onaka}, {Pickles},
  {Rhoads}, {Simon}, {Szalay}, {Szapudi}, {Tholen}, {Tonry}, {Waterson}, \&
  {Wick}}]{2002SPIE.4836..154K}
{Kaiser}, N., {Aussel}, H., {Burke}, B.~E., {et~al.} 2002, in Society of
  Photo-Optical Instrumentation Engineers (SPIE) Conference Series, Vol. 4836,
  Survey and Other Telescope Technologies and Discoveries, ed. J.~A. {Tyson} \&
  S.~{Wolff}, 154--164, \dodoi{10.1117/12.457365}

\bibitem[{{Kasen} \& {Bildsten}(2010)}]{2010ApJ...717..245K}
{Kasen}, D., \& {Bildsten}, L. 2010, \apj, 717, 245,
  \dodoi{10.1088/0004-637X/717/1/245}

\bibitem[{{Kasen} {et~al.}(2017){Kasen}, {Metzger}, {Barnes}, {Quataert}, \&
  {Ramirez-Ruiz}}]{2017Natur.551...80K}
{Kasen}, D., {Metzger}, B., {Barnes}, J., {Quataert}, E., \& {Ramirez-Ruiz}, E.
  2017, \nat, 551, 80, \dodoi{10.1038/nature24453}

\bibitem[{{Korobkin} {et~al.}(2012){Korobkin}, {Rosswog}, {Arcones}, \&
  {Winteler}}]{2012MNRAS.426.1940K}
{Korobkin}, O., {Rosswog}, S., {Arcones}, A., \& {Winteler}, C. 2012, \mnras,
  426, 1940, \dodoi{10.1111/j.1365-2966.2012.21859.x}

\bibitem[{{Lauberts} \& {Valentijn}(1989)}]{1989spce.book.....L}
{Lauberts}, A., \& {Valentijn}, E.~A. 1989, {The surface photometry catalogue
  of the ESO-Uppsala galaxies}

\bibitem[{{Li} \& {Paczy{\'n}ski}(1998)}]{1998ApJ...507L..59L}
{Li}, L.-X., \& {Paczy{\'n}ski}, B. 1998, \apjl, 507, L59,
  \dodoi{10.1086/311680}

\bibitem[{{Liu} {et~al.}(2023){Liu}, {Lin}, {Yu}, {Wang}, {Mourani}, {Zhao}, \&
  {Dai}}]{2023ApJ...947...59L}
{Liu}, Z.-Y., {Lin}, Z.-Y., {Yu}, J.-M., {et~al.} 2023, \apj, 947, 59,
  \dodoi{10.3847/1538-4357/acc73b}

\bibitem[{{L{\"u}} {et~al.}(2015){L{\"u}}, {Zhang}, {Lei}, {Li}, \&
  {Lasky}}]{2015ApJ...805...89L}
{L{\"u}}, H.-J., {Zhang}, B., {Lei}, W.-H., {Li}, Y., \& {Lasky}, P.~D. 2015,
  \apj, 805, 89, \dodoi{10.1088/0004-637X/805/2/89}

\bibitem[{{Ma} {et~al.}(2025){Ma}, {Wang}, {Mo}, {Andrew Howell}, {Pellegrino},
  {Zhang}, {Wu}, {Yan}, {Liu}, {Arcavi}, {Chen}, {Farah}, {Padilla Gonzalez},
  {Guo}, {Hiramatsu}, {Li}, {Lin}, {Liu}, {McCully}, {Newsome}, {Sai},
  {Terreran}, {Xiang}, \& {Zhang}}]{2025A&A...698A.306M}
{Ma}, X., {Wang}, X., {Mo}, J., {et~al.} 2025, \aap, 698, A306,
  \dodoi{10.1051/0004-6361/202452685}

\bibitem[{{Makhathini} {et~al.}(2021){Makhathini}, {Mooley}, {Brightman},
  {Hotokezaka}, {Nayana}, {Intema}, {Dobie}, {Lenc}, {Perley}, {Fremling},
  {Mold{\`o}n}, {Lazzati}, {Kaplan}, {Balasubramanian}, {Brown}, {Carbone},
  {Chandra}, {Corsi}, {Camilo}, {Deller}, {Frail}, {Murphy}, {Murphy}, {Nakar},
  {Smirnov}, {Beswick}, {Fender}, {Hallinan}, {Heywood}, {Kasliwal}, {Lee},
  {Lu}, {Rana}, {Perkins}, {White}, {J{\'o}zsa}, {Hugo}, \&
  {Kamphuis}}]{2021ApJ...922..154M}
{Makhathini}, S., {Mooley}, K.~P., {Brightman}, M., {et~al.} 2021, \apj, 922,
  154, \dodoi{10.3847/1538-4357/ac1ffc}

\bibitem[{{Margalit} {et~al.}(2022){Margalit}, {Jermyn}, {Metzger}, {Roberts},
  \& {Quataert}}]{2022ApJ...939...51M}
{Margalit}, B., {Jermyn}, A.~S., {Metzger}, B.~D., {Roberts}, L.~F., \&
  {Quataert}, E. 2022, \apj, 939, 51, \dodoi{10.3847/1538-4357/ac8b01}

\bibitem[{{Metzger} {et~al.}(2008){Metzger}, {Quataert}, \&
  {Thompson}}]{2008MNRAS.385.1455M}
{Metzger}, B.~D., {Quataert}, E., \& {Thompson}, T.~A. 2008, \mnras, 385, 1455,
  \dodoi{10.1111/j.1365-2966.2008.12923.x}

\bibitem[{{Metzger} {et~al.}(2010){Metzger}, {Mart{\'\i}nez-Pinedo}, {Darbha},
  {Quataert}, {Arcones}, {Kasen}, {Thomas}, {Nugent}, {Panov}, \&
  {Zinner}}]{2010MNRAS.406.2650M}
{Metzger}, B.~D., {Mart{\'\i}nez-Pinedo}, G., {Darbha}, S., {et~al.} 2010,
  \mnras, 406, 2650, \dodoi{10.1111/j.1365-2966.2010.16864.x}

\bibitem[{{Mooley} {et~al.}(2018){Mooley}, {Deller}, {Gottlieb}, {Nakar},
  {Hallinan}, {Bourke}, {Frail}, {Horesh}, {Corsi}, \&
  {Hotokezaka}}]{2018Natur.561..355M}
{Mooley}, K.~P., {Deller}, A.~T., {Gottlieb}, O., {et~al.} 2018, \nat, 561,
  355, \dodoi{10.1038/s41586-018-0486-3}

\bibitem[{{Musolino} {et~al.}(2025){Musolino}, {Rezzolla}, \&
  {Most}}]{2025ApJ...984L..61M}
{Musolino}, C., {Rezzolla}, L., \& {Most}, E.~R. 2025, \apjl, 984, L61,
  \dodoi{10.3847/2041-8213/adcd6d}

\bibitem[{{Nicholl} {et~al.}(2017){Nicholl}, {Guillochon}, \&
  {Berger}}]{2017ApJ...850...55N}
{Nicholl}, M., {Guillochon}, J., \& {Berger}, E. 2017, \apj, 850, 55,
  \dodoi{10.3847/1538-4357/aa9334}

\bibitem[{{Niu} {et~al.}(2025){Niu}, {Hanna}, {Haster}, {Adhicary}, {Baral},
  {Baylor}, {Cousins}, {Creighton}, {Fong}, {Huang}, {Huxford}, {Joshi},
  {Kennington}, {Li}, {Magee}, {Meacher}, {Messick}, {Morisaki}, {Posnansky},
  {Sachdev}, {Sakon}, {Shah}, {Singh}, {Tapia}, {Tsukada}, {Viets},
  {Yarbrough}, \& {Zhang}}]{Niu25}
{Niu}, W., {Hanna}, C., {Haster}, C.-J., {et~al.} 2025, arXiv e-prints,
  arXiv:2509.09741.
\newblock \doarXiv{2509.09741}

\bibitem[{{{\"O}zel} \& {Freire}(2016)}]{2016ARA&A..54..401O}
{{\"O}zel}, F., \& {Freire}, P. 2016, \araa, 54, 401,
  \dodoi{10.1146/annurev-astro-081915-023322}

\bibitem[{{Paturel} {et~al.}(2003){Paturel}, {Petit}, {Prugniel}, {Theureau},
  {Rousseau}, {Brouty}, {Dubois}, \& {Cambr{\'e}sy}}]{2003A&A...412...45P}
{Paturel}, G., {Petit}, C., {Prugniel}, P., {et~al.} 2003, \aap, 412, 45,
  \dodoi{10.1051/0004-6361:20031411}

\bibitem[{{Piro} {et~al.}(2017){Piro}, {Giacomazzo}, \&
  {Perna}}]{2017ApJ...844L..19P}
{Piro}, A.~L., {Giacomazzo}, B., \& {Perna}, R. 2017, \apjl, 844, L19,
  \dodoi{10.3847/2041-8213/aa7f2f}

\bibitem[{{Planck Collaboration} {et~al.}(2020){Planck Collaboration},
  {Aghanim}, {Akrami}, {Ashdown}, {Aumont}, {Baccigalupi}, {Ballardini},
  {Banday}, {Barreiro}, {Bartolo}, {Basak}, {Battye}, {Benabed}, {Bernard},
  {Bersanelli}, {Bielewicz}, {Bock}, {Bond}, {Borrill}, {Bouchet}, {Boulanger},
  {Bucher}, {Burigana}, {Butler}, {Calabrese}, {Cardoso}, {Carron},
  {Challinor}, {Chiang}, {Chluba}, {Colombo}, {Combet}, {Contreras}, {Crill},
  {Cuttaia}, {de Bernardis}, {de Zotti}, {Delabrouille}, {Delouis}, {Di
  Valentino}, {Diego}, {Dor{\'e}}, {Douspis}, {Ducout}, {Dupac}, {Dusini},
  {Efstathiou}, {Elsner}, {En{\ss}lin}, {Eriksen}, {Fantaye}, {Farhang},
  {Fergusson}, {Fernandez-Cobos}, {Finelli}, {Forastieri}, {Frailis},
  {Fraisse}, {Franceschi}, {Frolov}, {Galeotta}, {Galli}, {Ganga},
  {G{\'e}nova-Santos}, {Gerbino}, {Ghosh}, {Gonz{\'a}lez-Nuevo}, {G{\'o}rski},
  {Gratton}, {Gruppuso}, {Gudmundsson}, {Hamann}, {Handley}, {Hansen},
  {Herranz}, {Hildebrandt}, {Hivon}, {Huang}, {Jaffe}, {Jones}, {Karakci},
  {Keih{\"a}nen}, {Keskitalo}, {Kiiveri}, {Kim}, {Kisner}, {Knox},
  {Krachmalnicoff}, {Kunz}, {Kurki-Suonio}, {Lagache}, {Lamarre}, {Lasenby},
  {Lattanzi}, {Lawrence}, {Le Jeune}, {Lemos}, {Lesgourgues}, {Levrier},
  {Lewis}, {Liguori}, {Lilje}, {Lilley}, {Lindholm}, {L{\'o}pez-Caniego},
  {Lubin}, {Ma}, {Mac{\'\i}as-P{\'e}rez}, {Maggio}, {Maino}, {Mandolesi},
  {Mangilli}, {Marcos-Caballero}, {Maris}, {Martin}, {Martinelli},
  {Mart{\'\i}nez-Gonz{\'a}lez}, {Matarrese}, {Mauri}, {McEwen}, {Meinhold},
  {Melchiorri}, {Mennella}, {Migliaccio}, {Millea}, {Mitra},
  {Miville-Desch{\^e}nes}, {Molinari}, {Montier}, {Morgante}, {Moss}, {Natoli},
  {N{\o}rgaard-Nielsen}, {Pagano}, {Paoletti}, {Partridge}, {Patanchon},
  {Peiris}, {Perrotta}, {Pettorino}, {Piacentini}, {Polastri}, {Polenta},
  {Puget}, {Rachen}, {Reinecke}, {Remazeilles}, {Renzi}, {Rocha}, {Rosset},
  {Roudier}, {Rubi{\~n}o-Mart{\'\i}n}, {Ruiz-Granados}, {Salvati}, {Sandri},
  {Savelainen}, {Scott}, {Shellard}, {Sirignano}, {Sirri}, {Spencer},
  {Sunyaev}, {Suur-Uski}, {Tauber}, {Tavagnacco}, {Tenti}, {Toffolatti},
  {Tomasi}, {Trombetti}, {Valenziano}, {Valiviita}, {Van Tent}, {Vibert},
  {Vielva}, {Villa}, {Vittorio}, {Wandelt}, {Wehus}, {White}, {White},
  {Zacchei}, \& {Zonca}}]{2020A&A...641A...6P}
{Planck Collaboration}, {Aghanim}, N., {Akrami}, Y., {et~al.} 2020, \aap, 641,
  A6, \dodoi{10.1051/0004-6361/201833910}

\bibitem[{{Salafia} {et~al.}(2022){Salafia}, {Colombo}, {Gabrielli}, \&
  {Mandel}}]{2022A&A...666A.174S}
{Salafia}, O.~S., {Colombo}, A., {Gabrielli}, F., \& {Mandel}, I. 2022, \aap,
  666, A174, \dodoi{10.1051/0004-6361/202243260}

\bibitem[{{S{\'a}nchez} {et~al.}(2014){S{\'a}nchez}, {Carrasco Kind}, {Lin},
  {Miquel}, {Abdalla}, {Amara}, {Banerji}, {Bonnett}, {Brunner}, {Capozzi},
  {Carnero}, {Castander}, {da Costa}, {Cunha}, {Fausti}, {Gerdes}, {Greisel},
  {Gschwend}, {Hartley}, {Jouvel}, {Lahav}, {Lima}, {Maia}, {Mart{\'\i}},
  {Ogando}, {Ostrovski}, {Pellegrini}, {Rau}, {Sadeh}, {Seitz},
  {Sevilla-Noarbe}, {Sypniewski}, {de Vicente}, {Abbot}, {Allam}, {Atlee},
  {Bernstein}, {Bernstein}, {Buckley-Geer}, {Burke}, {Childress}, {Davis},
  {DePoy}, {Dey}, {Desai}, {Diehl}, {Doel}, {Estrada}, {Evrard},
  {Fern{\'a}ndez}, {Finley}, {Flaugher}, {Frieman}, {Gaztanaga}, {Glazebrook},
  {Honscheid}, {Kim}, {Kuehn}, {Kuropatkin}, {Lidman}, {Makler}, {Marshall},
  {Nichol}, {Roodman}, {S{\'a}nchez}, {Santiago}, {Sako}, {Scalzo}, {Smith},
  {Swanson}, {Tarle}, {Thomas}, {Tucker}, {Uddin}, {Vald{\'e}s}, {Walker},
  {Yuan}, \& {Zuntz}}]{2014MNRAS.445.1482S}
{S{\'a}nchez}, C., {Carrasco Kind}, M., {Lin}, H., {et~al.} 2014, \mnras, 445,
  1482, \dodoi{10.1093/mnras/stu1836}

\bibitem[{{Siegel} \& {Metzger}(2017)}]{2017PhRvL.119w1102S}
{Siegel}, D.~M., \& {Metzger}, B.~D. 2017, \prl, 119, 231102,
  \dodoi{10.1103/PhysRevLett.119.231102}

\bibitem[{{Smith} {et~al.}(2020){Smith}, {Smartt}, {Young}, {Tonry}, {Denneau},
  {Flewelling}, {Heinze}, {Weiland}, {Stalder}, {Rest}, {Stubbs}, {Anderson},
  {Chen}, {Clark}, {Do}, {F{\"o}rster}, {Fulton}, {Gillanders}, {McBrien},
  {O'Neill}, {Srivastav}, \& {Wright}}]{2020PASP..132h5002S}
{Smith}, K.~W., {Smartt}, S.~J., {Young}, D.~R., {et~al.} 2020, \pasp, 132,
  085002, \dodoi{10.1088/1538-3873/ab936e}

\bibitem[{Soni {et~al.}(2025)}]{LIGO:2024kkz}
Soni, S., {et~al.} 2025, Class. Quant. Grav., 42, 085016,
  \dodoi{10.1088/1361-6382/adc4b6}

\bibitem[{{Sun} {et~al.}(2023){Sun}, {Wang}, {Yang}, {Zhang}, {Xiong}, {Yin},
  {Liu}, {Li}, {Xue}, {Yan}, {Zhang}, {Tan}, {Pan}, {Liu}, {Cheng}, {Zhang},
  {Hu}, {Zheng}, {An}, {Cai}, {Hu}, {Jin}, {Li}, {Li}, {Liu}, {Liu}, {Peng},
  {Song}, {Sun}, {Sun}, {Wang}, {Wen}, {Xiao}, {Yi}, {Zhang}, {Zhang}, {Zhang},
  {Zhang}, {Zhao}, {Zheng}, {Ling}, {Zhang}, {Yuan}, \&
  {Zhang}}]{2023arXiv230705689S}
{Sun}, H., {Wang}, C.~W., {Yang}, J., {et~al.} 2023, arXiv e-prints,
  arXiv:2307.05689, \dodoi{10.48550/arXiv.2307.05689}

\bibitem[{Taddia {et~al.}(2015)Taddia, Sollerman, Leloudas, Stritzinger,
  Valenti, Galbany, Kessler, Schneider, \& Wheeler}]{Taddia_2015}
Taddia, F., Sollerman, J., Leloudas, G., {et~al.} 2015, Astronomy \&
  Astrophysics, 574, A60, \dodoi{10.1051/0004-6361/201423915}

\bibitem[{{Tonry} {et~al.}(2018){Tonry}, {Denneau}, {Heinze}, {Stalder},
  {Smith}, {Smartt}, {Stubbs}, {Weiland}, \& {Rest}}]{2018PASP..130f4505T}
{Tonry}, J.~L., {Denneau}, L., {Heinze}, A.~N., {et~al.} 2018, \pasp, 130,
  064505, \dodoi{10.1088/1538-3873/aabadf}

\bibitem[{{van Velzen} {et~al.}(2020){van Velzen}, {Holoien}, {Onori}, {Hung},
  \& {Arcavi}}]{2020SSRv..216..124V}
{van Velzen}, S., {Holoien}, T. W.~S., {Onori}, F., {Hung}, T., \& {Arcavi}, I.
  2020, \ssr, 216, 124, \dodoi{10.1007/s11214-020-00753-z}

\bibitem[{{Yu} {et~al.}(2013){Yu}, {Zhang}, \& {Gao}}]{2013ApJ...776L..40Y}
{Yu}, Y.-W., {Zhang}, B., \& {Gao}, H. 2013, \apjl, 776, L40,
  \dodoi{10.1088/2041-8205/776/2/L40}

\end{thebibliography}
\bibliographystyle{aasjournal}

\end{document}